\documentclass[preprint,journal]{vgtc}       

\ifpdf
  \pdfoutput=1\relax                   
  \usepackage{graphicx}                
  \DeclareGraphicsExtensions{.pdf,.png,.jpg,.jpeg} 
\else
  \usepackage{graphicx}                
  \DeclareGraphicsExtensions{.eps}     
\fi%

\graphicspath{{figures/}{pictures/}{images/}{./}} 

\usepackage{microtype}                 
\PassOptionsToPackage{warn}{textcomp}  
\usepackage{textcomp}                  
\usepackage{mathptmx}                  
\usepackage{times}                     
\usepackage{cite}                      
\usepackage{tabu}                      
\usepackage{booktabs}                  
\usepackage[colorlinks = true,
            linkcolor = black,
            urlcolor  = blue,
            citecolor= black]{hyperref} 

\usepackage{enumitem}
\usepackage{review}

\newcommand{\crit}[1]{{\fontfamily{qag}\selectfont{\small{#1}}}}

\usepackage{ifthen}
\makeatletter
\newcommand*{\trrrace}{\begingroup\@makeother\#\@trrrace}
\newcommand*{\@trrrace}[2][]{%
  \href{https://vdl.sci.utah.edu/trrrace/?view=timeline&type=doc&id=#2}{\ifthenelse{\equal{#1}{}}T{#2}{#2}}%
  \endgroup}
\makeatother

\usepackage{color}
\definecolor{cb_blue}{rgb}{0.22,0.49,0.72}
\definecolor{cb_green}{rgb}{0.3,0.67,0.29}
\definecolor{cb_orange}{rgb}{0.84, 0.45, 0.06}
\definecolor{cb_red}{rgb}{1.0, 0.0, 0.0}
\definecolor{cb_periwinkle}{rgb}{0.4, 0.2, 0.8}
\definecolor{black}{rgb}{0,0,0}

\preprinttext{To appear in IEEE Transactions on Visualization and Computer Graphics.}

\onlineid{0}

\vgtccategory{Research}
\vgtcpapertype{application/design study}

\title{Insights From Experiments With Rigor in an EvoBio Design Study}

\author{Jen Rogers, Austin H. Patton, Luke Harmon, Alexander Lex, Miriah Meyer}
\authorfooter{
\item
 Jen Rogers, Alexander Lex, Miriah Meyer are with the University of Utah, E-mail: jen@sci.utah.edu, alex@sci.utah.edu, miriah@cs.utah.edu.
\item
 Austin H. Patton is with Washington State University, E-mail: austin.patton@wsu.edu.
\item
 Luke Harmon is with the University of Idaho, E-mail: lukeh@uidaho.edu.

}

\shortauthortitle{Biv \MakeLowercase{\textit{et al.}}: Insights From Experiments With Rigor}

\abstract{
Design study is an established approach of conducting problem-driven visualization research. The academic visualization community has produced a large body of work for reporting on design studies, informed by a handful of theoretical frameworks, and applied to a broad range of application areas. The result is an abundance of reported insights into visualization design, with an emphasis on novel visualization techniques and systems as the primary contribution of these studies. In recent work we proposed a new, interpretivist perspective on design study and six companion criteria for rigor that highlight the opportunities for researchers to contribute knowledge that extends beyond visualization idioms and software. In this work we conducted a year-long collaboration with evolutionary biologists to develop an interactive tool for visual exploration of multivariate datasets and phylogenetic trees. During this design study we experimented with methods to support three of the rigor criteria: \crit{ABUNDANT}, \crit{REFLEXIVE}, and \crit{TRANSPARENT}. As a result we contribute two novel visualization techniques for the analysis of multivariate phylogenetic datasets, three methodological recommendations for conducting design studies drawn from reflections over our process of experimentation, and two writing devices for reporting interpretivist design study. We offer this work as an example for implementing the rigor criteria to produce a diverse range of knowledge contributions.
} 


\CCScatlist{ 
 \CCScat{K.6.1}{Management of Computing and Information Systems}%
{Project and People Management}{Life Cycle};
 \CCScat{K.7.m}{The Computing Profession}{Miscellaneous}{Ethics}
}

\renewcommand{\manuscriptnotetxt}{This is the authors’ preprint version of this paper. Please
cite the following reference:
Jen Rogers, Austin H. Patton, Luke Harmon, Alexander Lex, Miriah Meyer. Insights From Experiments With Rigor in an EvoBio Design Study. IEEE Transactions on Visualization and Computer Graphics,
to appear, 2020.}

\begin{document}
\firstsection{Introduction}
\maketitle

Design study is an established approach to problem-driven visualization inquiry that emphasizes designing visual analysis tools in close collaboration with domain experts \cite{sedlmair_design_2012}.
Within a design study, visualization researchers build an understanding of a problem domain and translate that understanding into a visualization design, iteratively refining both their understanding of the problem and their visual analysis solution through close work with domain collaborators. 
Researchers conducting design studies draw from a host of theoretical constructs to guide the inquiry process, from process models \cite{sedlmair_design_2012,mckenna_design_2014,mccurdy_action_2016,meyer_criteria_2019,hall_design_2019} to design decision models \cite{munzner_nested_2009,meyer2015nested}, guiding scenarios \cite{sedlmair2016design}, educational models \cite{syeda_design_2020}, and collaboration roles \cite{simon2015bridging,wong2018towards}. As a result, an increasing number of reports describe effective design studies within a broad range of application areas \cite{nobre_lineage:_2019, partl_pathfinder:_2016, mccurdy2015poemage, zhang2018idmvis, law2018duet, hinrichs2015speculative, brehmer2014overview,mccurdy_framework_2019}.

Historically, design study papers have emphasized novel visual analysis systems and techniques as primary knowledge contributions \cite{meyer_criteria_2019}. Many of these papers also cite domain characterizations and abstractions \cite{munzner_nested_2009} as contributions under the reasoning that they are important for judging the validity of technical design artifacts and for building a body of visual analysis requirements that others can design against. The original definition of design study also includes lessons-learned as a potential knowledge contribution stemming from reflection, but scant guidance is available on how to generate knowledge of this sort \cite{2018_cga_reflection}.

In Meyer \& Dykes \cite{meyer_criteria_2019} we proposed a new, interpretivist view of visualization design study to produce a more diverse range of knowledge contributions. As a critique of the software-centric view of design study, this new perspective emphasizes the potential for using design study to acquire a more diverse range of knowledge, including knowledge about the visualization design process as well as about people's relationship with data and technology more broadly. This work recommends six rigor criteria for guiding the design study process toward acquiring new knowledge:  
\crit{INFORMED}, \crit{REFLEXIVE}, \crit{ABUNDANT}, \crit{PLAUSIBLE}, \crit{RESONANT}, and \crit{TRANSPARENT}.
These criteria provide an opportunity for researchers to rethink how to conduct effective design studies, learning new things along the way.

In the work we present here we experimented with methods to support three of the rigor criteria:
\crit{ABUNDANT}, \crit{REFLEXIVE}, and \crit{TRANSPARENT}. Our experimentations took place within the context of a one-year design study with evolutionary biologists. We employed techniques such as an immersive, three-month field study; structured and systematic reflection; and careful curation of documents and other design artifacts. Through a period of collaborative, critical reflection, we identified several methodological insights that emerged from our experiments. 

The resulting contributions from this inquiry are diverse, including both technical and methodological insights. 
More specifically, the contributions include:
\begin{itemize}[nolistsep,noitemsep]
    \item Two new visualization techniques for supporting the analysis of multivariate trees: (1) a \textit{trait view} that visualizes node-value distributions under uncertainty for associated characteristics along multivariate subtrees; and (2) a \textit{pattern view} that aids in the discovery and visualization of patterns in value trajectories for attributes across paths in a tree.
    \item Three methodological recommendations for conducting interpretivist design study: (1) establish systematic reflective practices that include reflexive notes, reflective transcriptions, and artifact curation; (2) build and maintain a trace of diverse research artifacts; and (3) argue for rigor from evidence, not just methods.
    \item Two experimental writing devices for reporting on interpretivist design study: (1) inclusion of direct links to research artifacts to transparently provide an abundance of evidence; and (2) embedding of a design study paper within a methodological one to highlight the diversity of our research contributions.
\end{itemize}
This work serves as an example of how researchers can consider the \crit{ABUNDANT}, \crit{REFLEXIVE}, and \crit{TRANSPARENT} criteria in practice, as well as the diverse types of knowledge contributions possible through their consideration.

We first provide the theoretical backdrop for our methodological work in Section \ref{sec:theoretical-backdrop}, followed by a description of our research methods in Section \ref{sec:methods}. Section \ref{sec:trevo} is a design study paper-within-a-paper, emphasizing the technical aspect of this work; our methodological recommendations follow in Section \ref{sec:lessons-learned}. 
Throughout the paper we include direct links to our abundant collection of research artifacts --- for example [\trrrace{45}] ---  to transparently provide evidence for our claims.

\section{Theoretical Backdrop}
\label{sec:theoretical-backdrop}

The methodological work we present in this paper draws from the interpretivist perspective of design study proposed by Meyer \& Dykes \cite{meyer_criteria_2019}. This perspective argues for a myriad of opportunities for researchers to make valuable knowledge contributions beyond visualization techniques and software. Doing so, however, requires a rethinking of design study research practices and the ways we make quality judgments about the inquiry. Six criteria for rigor guide an interpretivist design study approach --- \crit{INFORMED}, \crit{REFLEXIVE}, \crit{ABUNDANT}, \crit{PLAUSIBLE}, \crit{RESONANT}, and \crit{TRANSPARENT} --- which are derived from theoretical positions in social science \cite{tracy2010qualitative,lincoln_establishing_1982,smith2018developing},
information systems \cite{sein2011action}, and research through design \cite{frayling1994research,zimmerman2007research}. Achieving all six criteria within a single design study is unlikely to occur due to pragmatic constraints such as time and resources \cite{meyer_criteria_2019}. In the work presented in this paper, we focus on \crit{ABUNDANT}, \crit{REFLEXIVE}, and \crit{TRANSPARENT}, exploring various ways to achieve these criteria, as well as the kinds of knowledge elucidated by doing so.

\subsection{Abundant}
\label{sec:abundance}

A design study with abundance reflects the richness and complexity of the situation under study \cite{meyer_criteria_2019}. An abundant design study thus includes a rich and diverse body of evidence, as well as an abundance of other considerations such as participant voices, designs, and time in the field. In our experiments we considered all of these aspects of abundance.

The inclusion of a variety of voices and contexts reflects a valuing of \textit{pluralism} found in critical feminist theory that ``insists that the most complete knowledge comes from synthesizing multiple perspectives'' \cite{klein_data_2020}. In human computer interaction (HCI), pluralism is argued as a mechanism for resisting designs that embed ``any single, totalizing, or universal point of view'' \cite{bardzell2010feminist}. Arguments for pluralism can be grounded in the idea of situated knowledges \cite{haraway1988situated}, which argues an epistemic view of a singular reality that can only be known only partially, embedded within a specific context. It is by combining these partial perspectives --- through ``actively and deliberately inviting other perspectives into the data analysis'' \cite{klein_data_2020} --- that a researcher achieves a fuller, richer view of the situation under study. 

An emphasis on exploring a design space through many, rapid designs similarly helps a designer avoid blind spots and fixation on a singular solution \cite{buxton_sketching_2010,cross2004expertise}. Design problems are wicked by nature, with an extensive space of possible solutions \cite{buchanan_wicked_1992}.
By broadly considering a design space, designers are more likely to find good solutions, rather than average or poor ones \cite{sedlmair_design_2012}, as well as to develop a better understanding of the problem under study \cite{cross2004expertise}.  
Dow et al.\ recommend exploring and refining design ideas in parallel, rather then through a sequential process, to obtain better and more diverse design artifacts \cite{dow_parallel_2011}. In the same vein, Buxton advocates for rapid sketching with broad ideation for developing effective design concepts through iterations of ``controlled convergence'' \cite{buxton_sketching_2010}. 

Finally, abundance through prolonged engagement with the people and context under study is a mainstay of qualitative research \cite{lincoln_establishing_1982,shenton2004strategies,tracy2010qualitative}. Researchers who establish an early familiarity with a domain build trust with their participants as well as the ability to understand domain-specific nuances of what they observe: ``objects and behaviors take not only their meaning but their very existence from their contexts'' \cite{lincoln_establishing_1982}. In a visualization study, \textit{design by immersion} is an approach for engagement in which both the visualization researchers and domain experts ``participate in the work of another domain such that visualization design, solutions, and knowledge emerge from these transdisciplinary experiences and interactions'' \cite{hall_design_2019}. This methodology allows visualization researchers to enrich their understanding of a domain, explore a broader visualization solution space, and build trust and agency with collaborators. Field studies --- in which a researcher spends sustained time with participants in their natural environment --- is a technique that can support visualization researchers in achieving immersion through prolonged engagement \cite{mccurdy_framework_2019}.

\subsection{Reflexive}
\label{sec:reflexivity}

Being reflexive within a visualization design study is to strive for ``explicit and thoughtful self-awareness of a researcher's own role in a study'' \cite{meyer_criteria_2019}. As a cornerstone of interpretivist, qualitative research, reflexivity is an acknowledgement of a researcher's influence on a study, and vice versa \cite{birks_memoing_2008}. Researcher bias and perspective are an inherent part of qualitative research, and eliminating them from the research process is arguably impossible \cite{mantzoukas_inclusion_2005}. Reflexivity is instead an opportunity to gather valuable data \cite{rode_reflexivity_2011} that can help researchers understand their biases and perspectives as a vector for change and learning \cite{finlay2002negotiating}.

Reflexivity is an important consideration in the third wave of HCI research \cite{bodker2006second}. Largely discussed in the critical HCI literature, reflexivity is considered a mechanism for researchers ``to be accountable for the ways in which HCI construes design(ing) and acknowledge our responsibility \ldots~to challenge the dominant view on design'' \cite{avle_design(ing)_2016}.
Despite its importance, the HCI community has been slow to broadly adopt reflexive practices in research due to the scrutiny on subjectivity during the review process.
The visualization research community shares a similar emphasis and valuing of objectivity \cite{meyer_criteria_2019}, and a lack of methods for supporting and exploiting reflexivity. This gap motivated our experimentations with reflexivity.

Reflexivity is a type of (self) reflection \cite{macbeth_"reflexivity"_2016}.
As a method, reflection traces to Sch\"{o}n's ideas of reflective practice through reflection-in-action and reflection-on-action \cite{schon_reflective_1984}. 
Reflection-in-action is characterized as an intuitive, rapid, reflective response ``in the moment'' \cite{yanow_what_2009}. 
Reflection-on-action instead happens after an experience, and is characterized as an ``inquiry into the personal theories that lie as the basis of one's actions'' \cite{korthagen_linking_2001}.  
A commonly employed method for reflection-on-action in qualitative research is \textit{memoing}: ``Memos can help to clarify thinking on a research topic, provide a mechanism for the articulation of assumptions and subjective perspectives about the area of research, and facilitate the development of the study design'' \cite{birks_memoing_2008}. We used memoing throughout our design study to facilitate reflexivity and reflection.

Pragmatically, reflection-on-action is synonymous with critical reflection \cite{daley_reflections_2010}, an inquiry process where researchers question their assumptions by examining the reasoning and ideology that frame their practice and experiences \cite{brookfield_critically_1998,thompson2018critically}. Work by Kerzner et al.\ employs critical reflection to construct a general framework for visualization workshops from their experiences running 17 of them \cite{kerzner_framework_2019}. Similarly, Satyanarayan et al.\ create a set of lessons for designing visualization authoring toolkits using what they call critical reflections \cite{satyanarayan2019critical}. Although not grounded in the reflection literature, their process is similar to that of reflection-on-action practices. 
Other than a handful of examples like these, the visualization literature is largely lacking pragmatic guidance on how and when to reflect \cite{2018_cga_reflection}; this work contributes actionable recommendations for reflecting in a design study.

\subsection{Transparent}
\label{sec:transparency}

Transparent reporting of a design study --- through scrutinizable documentation of data, methods, analysis, and artifacts --- is necessary for supporting judgments about the quality of the study and its results \cite{meyer_criteria_2019}. 
How to report transparently, however, is an open question. Recent work by Wacharamanotham et al.\ provides recommendations for sharing HCI research materials based on a survey of researchers \cite{wacharamanotham_transparency_2020}. This work, however, considers only software and hardware prototypes for design-oriented studies, missing many of the diverse artifacts produced within a design study such as sketches, abstractions, reflexive notes, and diagrams. In this work we experimented with recording and reporting a diverse set of design artifacts, drawing from ideas in qualitative research and research through design.

In interpretivist, qualitative research, the \textit{audit trail} is an established mechanism for transparent reporting \cite{lincoln_establishing_1982,akkerman2008auditing,de2018reflections,carcary2009research}. An audit trail is a detailed documentation of a research process that is intended for use in an \textit{audit process} \cite{akkerman2008auditing}. This process is undertaken by an (external) auditor who reviews the audit trail in order to asses the quality of the study, enhancing the trustworthiness of the research \cite{lincoln_establishing_1982}. Although audit trails are meant to increase the transparency of a study, they can also increase the quality through explicit thoughtfulness on the part of the researcher on what and how to record \cite{de2018reflections}. Two recent visualization design studies include audit trails as supplemental materials \cite{kerzner_framework_2019,mccurdy_framework_2019}, but neither study performed an audit. 

Transparently reporting on design decisions and insights is challenging due to the ingrained nature of knowledge within the artifacts themselves.
Design scholars consider the knowledge that a designer acquires to reside in the artifacts they create \cite{cross1999design}. This knowledge, however, is implicit and often opaque \cite{stappers2007doing}. \textit{Annotated portfolios} --- textual annotations of design patterns across a curated collection of designs --- is a method used within the research-through-design community to explicitly communicate knowledge embedded within designs \cite{gaverbill_annotated_2012,bowers_logic_2012}.
Annotations allow for comparison of designs and highlight relationships between disparate works, from which designers can develop and communicate generalized, intermediate knowledge. A different approach to externalizing design knowledge is that of literate visualization, which engages the designer in reflective documentation during the creation of digital, visualization artifacts \cite{wood2018design}.

\section{Methods}
\label{sec:methods}

To explore how an interpretivist approach to design study changes what and how we learn, we set out with the goal of experimenting with three criteria --- \crit{ABUNDANT}, \crit{REFLEXIVE}, and \crit{TRANSPARENT} --- during an evolutionary biology design study. We positioned this work within the perspective that design studies are wicked, subjective, and diverse \cite{meyer_criteria_2019}. Rogers conducted a three-month, immersive field study, followed by a design phase and a reflection phase in collaboration with Lex and Meyer. 
In this section we provide details about our research site and domain collaborators, the ways we experimented with the criteria, and the methods we employed for data collection and analysis. We directly link to our abundant collection of evidence --- for example [\trrrace{45}] --- to provide transparent reporting of our process.

\subsection{Research Site and Participants}
\label{sec:sites-and-team}

Our study took place at two sites. In the first phase, we undertook a three-month field study in the Harmon Lab at the University of Idaho, which studies ecology and evolution through phylogenetic analysis. During this time, Rogers spent work-hours within the group's lab, immersed in conditions similar to those in which the evolutionary biology graduate students worked. The lab environment was open and social, with six desks spaced around the edges of the room, a community couch often inhabited by other graduate students who stopped by, and a white board filled with scattered drawings and notes. The graduate students used this space for their computational work, which was often analysis of the phylogenetic data and field sample measurements taken from summer field work. This lab was chosen based on a relationship established through a federally funded research project~\cite{multinet} between the Harmon Lab and the Visualization Design Lab at the University of Utah. The design and reflection phases took place within the Visualization Design Lab. 

During the field study we worked with seven evolutionary biology collaborators. Two primary collaborators during this phase were Harmon, the PI of the evolutionary biology lab, and Patton, a graduate student at Washington State University who works closely with the Harmon lab, often on-site. Both primary collaborators are co-authors on this paper. Five other graduate students in the lab served as secondary collaborators. All collaborators were involved with the interviews and informal feedback. The primary collaborators were additionally involved with the design and evaluation of our visualization techniques.  

\subsection{Criteria Considerations}

Our decision to focus on the \crit{ABUNDANT}, \crit{REFLEXIVE}, and \crit{TRANSPARENT} criteria stemmed from our experiences in previous studies and considerations of actionability [\trrrace{160}]. In previous work we attempted to instill transparency through collecting artifacts and releasing audit trails \cite{kerzner_framework_2019,mccurdy_framework_2019}. These experiences led to numerous conversations within our research group about how to record and report artifacts in design studies and other qualitative research studies. We saw this design study as an opportunity to systematically experiment with abundant recording and transparent reporting of evidence from the very start of a study. We included reflexivity based on the interests of the research team and the actionability of reflexive memoing. Our approaches to meeting these criteria evolved over the course of the study.

We attempted to instill abundance in our design study in four ways. First, 
we meticulously curated a rich collection of artifacts generated throughout the design study including field notes and reflective memos [\trrrace{48}], email correspondence [\trrrace{90}], sketchbook scans [\trrrace{81}], photos of collaborator sketches [\trrrace{55}], links to papers [\trrrace{87}], low- and high-fidelity visualization prototypes [\trrrace{158}, \trrrace{96}], and notes reflectively transcribed from audio recordings of meetings [\trrrace{36}]. 
Second, we conducted an immersive field study, in which Rogers situated herself as a peer in the Harmon Lab for three months. Working in the communal space of our domain collaborators, Rogers actively engaged in research meetings and reading clubs focused on evolutionary topics of interest. She learned how to use the analysis pipelines of her collaborators to get a deeper understanding of the domain problem space [\trrrace{47}, \trrrace{50}]. Through time, she gained a deeper understanding of the domain research and developed a personal investment in our collaborators' research and social dynamics. These activities encompass the \textit{communal}, \textit{personal}, and \textit{active} themes of immersive studies \cite{hall_design_2019}.
Third, we contacted domain experts outside the Harmon Lab in an attempt to include multiple voices and datasets. We sent emails to colleagues of the Harmon Lab, as well as evolutionary biology researchers at the University of Utah, inviting them to participate in the evaluation of our visualization designs [\trrrace{109}]. 
Fourth, we relied heavily on sketching to facilitate brainstorming of visualization ideas [\trrrace{43}, \trrrace{52}], 
to understand the domain space [\trrrace{10}, \trrrace{38}], to communicate with domain collaborators [\trrrace{55}], and to aid in reflective analysis [\trrrace{138}]. 

We implemented reflexivity during the field study through regular, reflective memoing by Rogers. These reflections were reflexive in nature and included documenting her feelings as she became more incorporated into the lab, her insecurities that were potentially limiting the research [\trrrace{3}, \trrrace{20}], her interpretations on social dynamics and friendships within the lab, and how those dynamics affected the research [\trrrace{18}]. Memoing was done before and after meetings and during pivot-point moments in the research process. 

In an attempt to transparently communicate the design study process, we created an auditable website from our collection of research artifacts, which is available at \url{http://vdl.sci.utah.edu/trrrace/}. This website which we call a \textit{trrrace} and discuss in more detail in Section \ref{sec:trrrace}, traces the project from the field study through the design and reflection phases, organizing the abundant collection of artifacts we recorded throughout. The artifacts are organized in an interactive timeline and are discoverable via annotations, descriptive metadata, and directly in the timeline.

\subsection{Data Collection}
\label{sec:data-collection}

We kept a meticulous collection of all recorded artifacts starting from the beginning of the field study in an attempt to record an abundance of evidence from our design study process and support transparency. These artifacts were generated throughout all three phases of research, but the content creation was concentrated during times of immersion in the field study, as well as during times of correspondence with collaborators in the design phase of the tool. Throughout the field study, Rogers interviewed members of the lab, taking reflective notes before and after every interview. 
Pre-interview reflections included a review of previous meeting notes and outlining an agenda [\trrrace{8}], and post-interview reflections summarized the main talking points and speculated about productive next steps [\trrrace{20}]. 
Additionally, she audio-recorded these interviews and reflectively transcribed \cite{mccurdy_framework_2019} them to capture the context of what was said when, how things were said, and her interpretation of the conversations [\trrrace{53}]. To capture a rich view the interviews, Rogers recorded any white-board diagrams [\trrrace{94}], scribbles [\trrrace{41}], or sketches [\trrrace{55}] that were generated during discussions. In addition to the pre- and post-interview reflections, Rogers also regularly wrote reflexive memos that included her feelings on her immersion in the lab, her insecurities that were possibly limiting the research, friendships, social dynamics, and how those dynamics affected the research [\trrrace{3}, \trrrace{18}, \trrrace{20}]. 

During the second week of the field study, Rogers conducted a creative visualization opportunities workshop \cite{kerzner_framework_2019} with the lab members to brainstorm about potential visualization directions. We took photos of all the materials generated from the workshop exercises and audio recorded the workshop 
[\trrrace{23},\trrrace{24},\trrrace{25},\trrrace{26},\trrrace{27},\trrrace{28},
\trrrace{29},\trrrace{30},\trrrace{31},\trrrace{32},\trrrace{33},\trrrace{34}
\trrrace{35},\trrrace{36}].

The beginning stages of sketching and prototyping began during the field study, but the bulk of the design work and tool development happened during the design phase. 
Our primary collaborators remained extensively involved in providing feedback on design iterations, with much of this feedback happening through video calls, email, and in two, short, subsequent visits to the Harmon Lab. We recorded feedback emails [\trrrace{90}, \trrrace{118}], notes from the in-person feedback sessions [\trrrace{125}], and memos capturing personal interpretations of the feedback [\trrrace{126}]. Design artifacts generated during this process include sketches [\trrrace{43}, \trrrace{45}, \trrrace{52}], mock-ups [\trrrace{59}], and screen-shots of prototype iterations [\trrrace{67}, \trrrace{73}, \trrrace{92}].

\subsection{Analysis}
\label{sec:analysis}

Analysis occurred during the final, reflective stage of the study when we started the construction of an audit trail as a website for collecting and annotating our diverse set of research artifacts. The website was initially designed to communicate the design study process with a high-level of transparency and detail. The  organization and curation of artifacts, however, became a powerful catalyst for reflection that led to significant methodological insights about our design process, as well as new directions for the design of the visualization tool.
Through collaborative, critical reflection among the visualization research team members, we iteratively developed a set of actionable recommendations for conducting interpretivist design study from our insights looking across the collection of artifacts.

\section{Trevo: An Evolutionary Biology Design Study}
\label{sec:trevo}

This design study was motivated by the complexity of our collaborators' problem in representing the rich, multivariate, and uncertain data in their analysis. 
They work extensively with trees that represent hypothesized explanations for how species are related.
In this design study we developed a web-based visualization tool Trevo, that allows them to analyze these trees with multivariate and uncertain attributes. 

We report on this design study in an abbreviated form as a paper-within-a-paper as part of our larger goal of highlighting the diverse contributions possible from interpretivist design study.
This experimental format emerged from our dissatisfaction with 
early paper drafts that followed a more traditional design study reporting structure [\trrrace{144}, \trrrace{159}]. We felt the traditional structure overly accentuated technical contributions while leaving little room for significant methodological discussions. We developed the paper-within-a-paper style to stress the role of the design study as a method of inquiry~\cite{meyer_criteria_2019} that reflects and reports on a more diverse type of knowledge.

\subsection{Biological Background}

The driving question in the field of evolutionary biology is 
why the living world evolved the way it did? To answer this question, researchers need to determine when a given trait evolved, such as a lizard's long tail, and whether a particular species possesses that trait as a result of common ancestry or of other forces such as the environment. 
To answer these questions, evolutionary biologists study a group of living organisms to establish hypotheses about evolutionary forces that can generalize to other species. For example, researchers study anole lizards to infer how environment influences evolution. Analysis begins in the field, where these researchers take samples of living species and measure their physical characteristics, such as a lizard's tail length, snout length, and body mass. They use these measurements of current species, typically along with DNA sequence data, to reconstruct physical characteristics of the ancestors in a species' phylogenetic history. These histories are then the basis of studying when and why traits evolved, and whether the physical characteristics of contemporary species are, or are not, a result of evolution from common ancestors.

Evolutionary relationships are commonly represented as a binary tree, referred to as a \textbf{phylogenetic tree}. These trees are usually reconstructed by modeling the evolution of a set of DNA sequences sampled from present-day species. The leaf nodes of the tree represent the contemporary species, whereas inner nodes represent their common ancestors. All nodes in the tree have associated characteristics described by a set of traits. Internal nodes (common ancestors) have estimated values for these traits. Leaf nodes (species) have measured values for traits.

A common structure evolutionary biologists work with is \textbf{clades}, which are subtrees of the larger phylogeny in which all species share a single, unique, common ancestor. For example, for anole lizards, the main clade of study is the genus \textit{Anolis}, a group of more than 400 species that all evolved from a common ancestral lizard. These subtrees are sometimes predefined, as is the case for well-established clades such as anoles, or they can be defined during analysis.

\begin{figure}[t] 
\includegraphics[width=\linewidth]{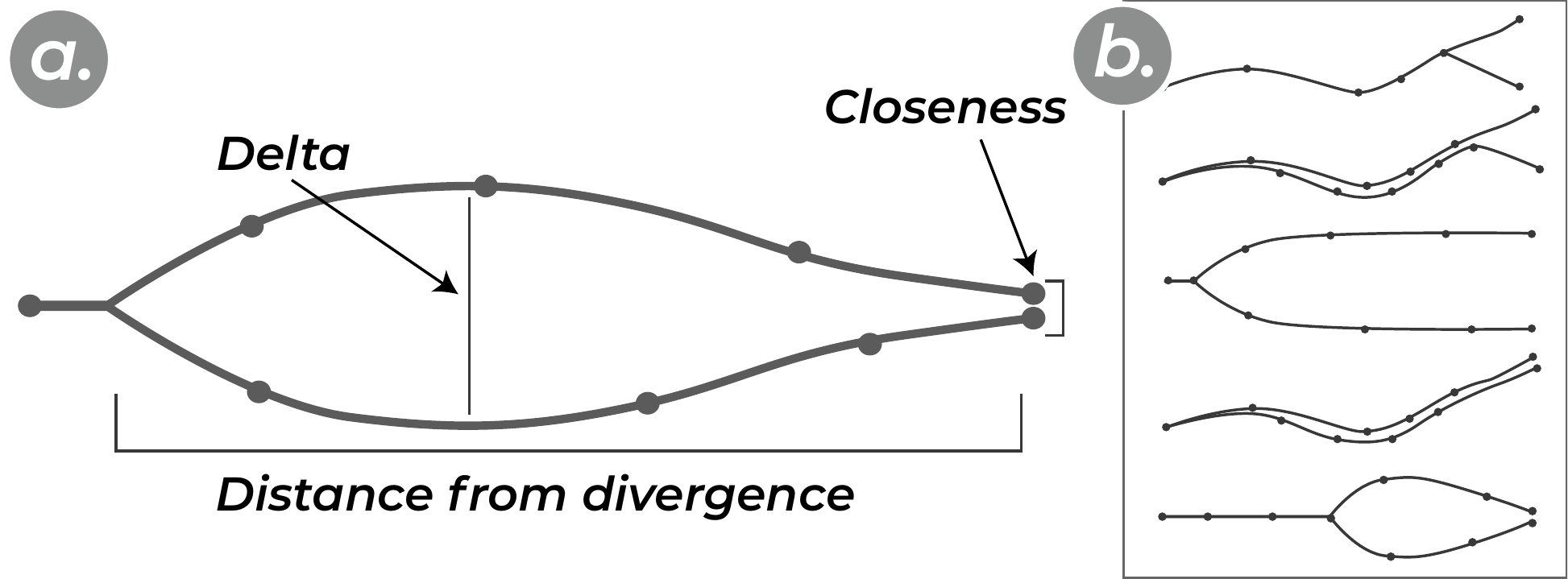}
 \vspace{-4mm}
 \caption{Defined preset patterns in the pattern view. (a) Pattern breakdown for convergence (b) The six predefined patterns.}
 \vspace{-7mm}
 \label{fig:pattern-breakdown}
\end{figure}

Researchers analyze different, possible evolutionary mechanisms by studying \textbf{patterns of evolution}. These patterns can be summarized in terms of how traits change, or evolve, along the branches of the phylogeny. A common pattern of trait evolution is that of \textbf{divergence} in which species evolve increasingly distinct trait values over time~\cite{rezende_phylogenetic_2012}. 
Another pattern, \textbf{convergence} which is shown in Figure~\ref{fig:pattern-breakdown}(a), is characterized by traits that diverge early in two species' histories, but then converge later in their evolutionary histories by developing similar or identical traits~\cite{ingram_surface:_2013}.
Convergence is an indication of adaption --- certain traits evolve repeatedly because they are beneficial in an environment --- and has been studied extensively in the anole lizards. Many of these lizards, having split off from their common ancestors a long time ago, inhabit similar environments on separated islands and have evolved very similar characteristics as a consequence. Although other interesting patterns besides divergence and convergence exist, such as those in Figure~\ref{fig:pattern-breakdown}(b), they do not have standardized names. 

Identifying patterns of evolution is a challenging analysis problem that involves accounting for changes to multiple traits under uncertainty in the context of the tree topology. We worked with our collaborators to explore new ways to enable this complex analysis with interactive visual analysis tools. 

\subsection{Data and Task Abstraction}

In the datasets our collaborators are analyzing, evolutionary relationships are represented as rooted trees. Bifurcations in the tree represent speciation events. Internal nodes encode hypothesized common ancestors of existing species, which in turn constitute the leaf nodes. The size of the trees we focused on here ranged from 20 to 200 species (leaves), each associated with 5 to 25 traits. Traits of a species can be discrete or continuous and are uncertain for the reconstructed  (inner node) species. Reconstructed discrete traits, such as the geographic location where a species is found or whether they lay eggs, are specified as probabilities. Continuous traits, such as tail length, are given as an estimated value and a 95\% confidence interval.

To explain why the living world evolved the way it did, our collaborators' analysis is focused on understanding when and how traits evolved in a population, which requires viewing trait values for multiple attributes in the context of the topology of the tree. We break down this larger analysis goal into three domain tasks:

\noindent\textbf{T1: Understand the uncertainty in multiple reconstructed traits.} \\
Significant uncertainty exists in the reconstructed traits for internal nodes, so adequate visual representations of trait values and their uncertainty are critical. Current methods for visualizing attributes in phylogenetic trees are limited to showing one or two traits at a time, and frequently cannot encode uncertainty [\trrrace{42}, \trrrace{36}, \trrrace{16}]. This task is orthogonal to all other tasks, i.e., uncertainty analysis is a part of every analysis task.

\noindent\textbf{T2: Analyze subtrees.}\\
This task is concerned with creating and analyzing individual subtrees (clades) and comparing between multiple subtrees.

\noindent\textbf{T2.1: Create subtrees.}\\
Our collaborators need the ability to create subtrees by topology and trait values. For example, an analyst might want to create two subtrees based on an attribute, such as the island a species is inhabiting [\trrrace{64}, \trrrace{80}]. Definitions of subtrees might also be given as formal clades in a dataset. 

\noindent\textbf{T2.2: Analyze attribute distributions in subtrees.}\\
Our collaborators need to be able to identify significant changes in multiple traits at once. For example, understanding whether a shift toward a longer tail is correlated with a shift toward longer hind-legs can give hints about the underlying causes of that change [\trrrace{20}]. Viewing multiple traits at once is particularly difficult for our collaborators, who rely on comparisons of reconstructed traits on separate trees [\trrrace{36}, \trrrace{72}]. 

\noindent\textbf{T2.3: Identify evolutionary outliers.}\\ 
It is important for our collaborators to identify individual species, paths, or subtrees that have significantly different trait values compared to the rest of the subtree [\trrrace{17}, \trrrace{91}]. For example, they want to identify paths with species that have a larger body mass than the rest of the subtree.

\noindent\textbf{T2.4: Compare attribute distributions of multiple subtrees.}\\ 
Comparisons are important in characterizing what makes a subtree unique. For example, our collaborators want to study whether the species in a subtree share common characteristics, such as head and tail length, that set them apart from the rest of the tree. To study how traits evolved through history, they need to understand how subtree trait distributions diverge and where this happens in the tree [\trrrace{4}, \trrrace{20}, \trrrace{66}, \trrrace{72}, \trrrace{88}]. 

\noindent\textbf{T3: Identify and analyze evolutionary patterns}\\
An important task in our collaborators' analysis is identifying the evolutionary patterns that indicate certain mechanisms underlying evolution [\trrrace{53}, \trrrace{64}, \trrrace{87}]. 
Identifying these patterns requires the comparison of trait trajectories of multiple species in a tree [\trrrace{80},\trrrace{93}]. To identify convergence, for example, an analyst would search for two paths that separated early in the tree with trait values that first diverged, but then later converged. 

\subsection{Related Work}

Visualization of phylogenetic data is challenging in three ways: (1) the trees can be large, requiring sophisticated navigation and/or aggregation strategies to browse them; (2) the topology of the trees is uncertain, requiring the comparison of multiple alternative trees; and (3) the trees are associated with many (uncertain) attributes, requiring sophisticated multivariate tree visualization strategies. Our work addresses the third problem, multivariate trees, but we briefly review all areas. 

The scale and uncertainty of topology remain challenges in phylogenetic research and numerous visual solutions have been proposed for both \cite{munzner_treejuxtaposer:_2003, lee_candidtree:_2007, liu_aggregated_2019, bouckaert_densitree:_2010, letunic_interactive_2019, rosindell_onezoom:_2012, block_deeptree_2012}. Large phylogenetic trees and topological uncertainty are not key problems for our collaborators; visualizing trees with many attributes, however, is.
As a generalization, visualizing many traits in the context of a tree is a type of multivariate network visualization problem. Nobre et al.~recently described the design space of a multivariate network visualization in a survey that included tree visualization~\cite{nobre_state_2019}. We here focus mostly on approaches for phylogenies but refer readers to this survey for a broader overview. 

Within the evolutionary biology community, visualizations of phylogenetic data are used for both exploration and presentation in papers. Most figures found in evolutionary biology papers show trees laid out using node-link diagrams with either linear or circular layouts, and on-node or on-edge encoding to show trait values~\cite{revell_two_2013, rezende_phylogenetic_2012}. These figures are often created with interactive tools such as iTOL~\cite{letunic_interactive_2019} or Dendroscope~\cite{huson_dendroscope_2012}, or using scripted plotting libraries, such as phytools or ggtree for R~\cite{revell_phytools:_2012, yu_using_2020}. Tools such as iTOL can visualize multiple attributes for the leaves, but the inner nodes are usually limited to a single attribute. Analysts, however, often need to account for multiple traits at once to identify underlying forces influencing trait change. In their current workflow, they compare different traits mapped to the nodes of multiple trees side-by-side. Such comparisons are difficult with just 2 traits, but analysts must often consider up to 10 traits for a given tree. As expressed by one of our collaborators, ``if you have 1 continuous trait you can do things. If you have 2 --- OK. If you have 3 or 4 or 5, there is nothing really sufficient'' [\trrrace{36}]. 

In the visualization community, several tools have been designed to visualize trees with attributes. Lineage~\cite{nobre_lineage:_2019}, for example, visualizes attributes for genealogical trees using a linearization approach, where the attributes are shown in a table; Juniper is a generalization of this method to networks~\cite{nobre_juniper:_2019}. Other tools, such as TreeVersity2~\cite{guerra-gomez_visualizing_2013}, visualize attributes using implicit layouts. Researchers currently have no tool suitable for visualizing many traits for inner nodes and leaves under uncertainty in the context of phylogenetic trees. 

\subsection{Visualization Design}
Two technical contributions emerged from this design study. The first is a technique for visualizing summary distributions of attributes in a (sub)tree --- the \textbf{trait view}--- designed to address the analysis of subtrees (T2). The second contribution is a view for querying, ranking, and visualizing patterns consisting of topological and attribute features --- the \textbf{pattern view} --- designed to address the identification and analysis of evolutionary patterns (T3). Both views visualize uncertainty (T1) and were implemented in a web-based tool we call Trevo, along with two additional views: \url{https://vdl.sci.utah.edu/Trevo/}.

\begin{figure*}[t]
\includegraphics[width=\linewidth]{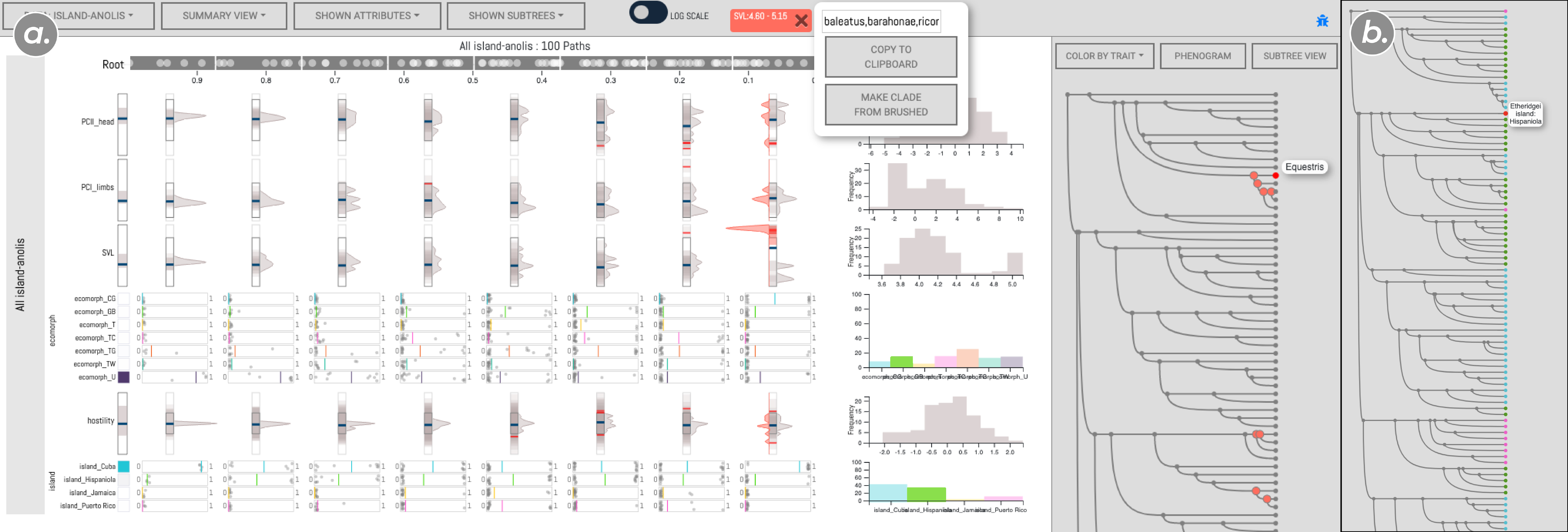}
 \vspace{-5mm}
 \caption{Trait view showing four continuous and two discrete trait variables for 100 \textit{Anolis} lizard species. (a) Outliers in the last SVL bin are brushed. A traditional phylogenetic tree view, shown on the right, can be used to define subtrees. (b) Leaf nodes can be color-coded by trait category. This detail view shows all leaf nodes color-coded by island of origin. These categories can be used to define subgroups by trait category or value, independent of the topology of the tree.} 
 \label{fig:trait_view}
  \vspace{-5mm}
\end{figure*}

\subsubsection{Trait View}

A crucial task for our collaborators is analyzing patterns of attributes within and between subtrees. When subtrees are defined topologically, this analysis can be supported in the context of a phylogenetic tree. For subtrees defined based on trait values, however, species can be scattered across a phylogentic tree. 
For example, our collaborators want to create two subtrees for anole species that are found on the islands of Hispaniola and Cuba so they can compare the distribution of body mass of the lizards on these islands to study any environmental effects that might appear. The ``island'' trait does not clearly split the phylogenetic tree into disjunct subtrees, as common ancestors colonized islands multiple times. It instead creates trees with partially overlapping branches. Figure-\ref{fig:trait_view}(b) shows these disjunct subtrees with the species color coded by island. Lizards originating from Hispaniola are colored green, and those originating from Cuba are colored blue. Our collaborators compare the subtrees' trait values through the evolutionary history to determine when and how these groups began to diverge, for example, to determine if there is a difference in body mass between the two islands and when this divergence in traits began to occur along the evolutionary history. Identifying differences in value trends and when they occur within the phylogenetic tree can be difficult given the overlapping topology.

\begin{figure}[t]
\includegraphics[width=\linewidth]{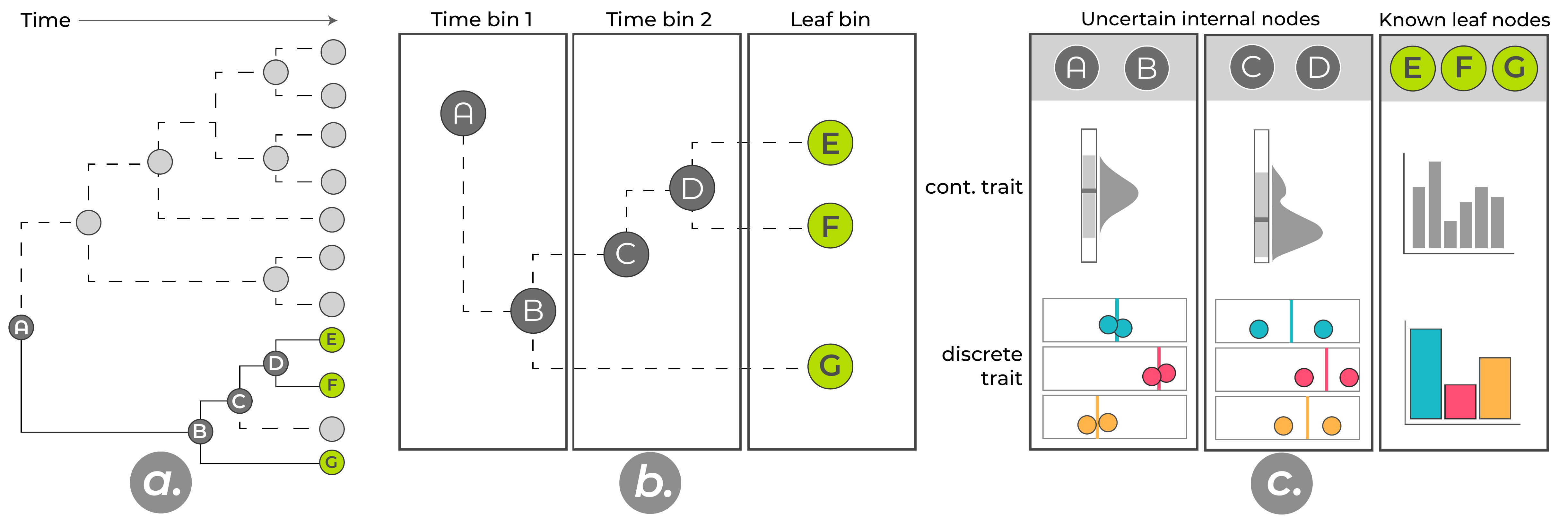}
 \vspace{-6mm}
 \caption{Transforming a phylogenetic tree into the trait view. (a) We select a subtree by brushing for a trait in the leaves, shown in green. (b) The subtree is binned by time intervals and the leaves are assigned a separate bin. (c) We show continuous uncertain traits using a median plus confidence interval visualization and a KDE plot. For discrete uncertain traits we use multiple dot-plots, one for each trait category. Known traits are visualized using histograms.}  \label{fig:trait-breakdown}
  \vspace{-6mm}
\end{figure}

Through an iterative design process with our primary collaborators [\trrrace{68}, \trrrace{74}, \trrrace{108}, \trrrace{114}], we tackled this challenge with an aggregation solution for creating trait-defined subtrees. The key aspect of this new trait view is that it enables analysts to filter branches of the tree based on traits of the leaves.
Figure~\ref{fig:trait-breakdown} shows the steps involved in transforming a node-link tree layout into the trait view. Initially, the tree is filtered to include only extant species with a certain attribute such as the green leaves in Figure~\ref{fig:trait-breakdown}(a). 
We then leverage temporal information to bin the other nodes in the subtree by time, shown in Figure~\ref{fig:trait-breakdown}(b). The leaves are assigned a separate bin for which the uncertain discrete- and numerical-trait distributions are visualized in columns. Nodes are shown at the top of the bin; their horizontal position is driven by their time attribute, allowing analysts to compare multiple uncertain trait distributions in a temporal context unhindered by the tree's topology.
Next, we use different encodings for leaf nodes with known trait values versus inner nodes with uncertain ones, shown in Figure~\ref{fig:trait-breakdown}(c). The known attributes of the leaves are encoded using histograms. For continuous uncertain traits we show the median plus a 95\% confidence interval for the estimated values and a kernel density estimate plot.
Finally, probabilities for uncertain discrete traits are represented in the trait view as separate one-dimensional dot plots for each state; 
to reduce the risk of overlapping dots, we use transparency and vertical jitter. 
The average for each state probability is plotted as a line in the plots.

\subsubsection{Pattern View}

\begin{figure*}[t]
\centering
\includegraphics[width=\textwidth]{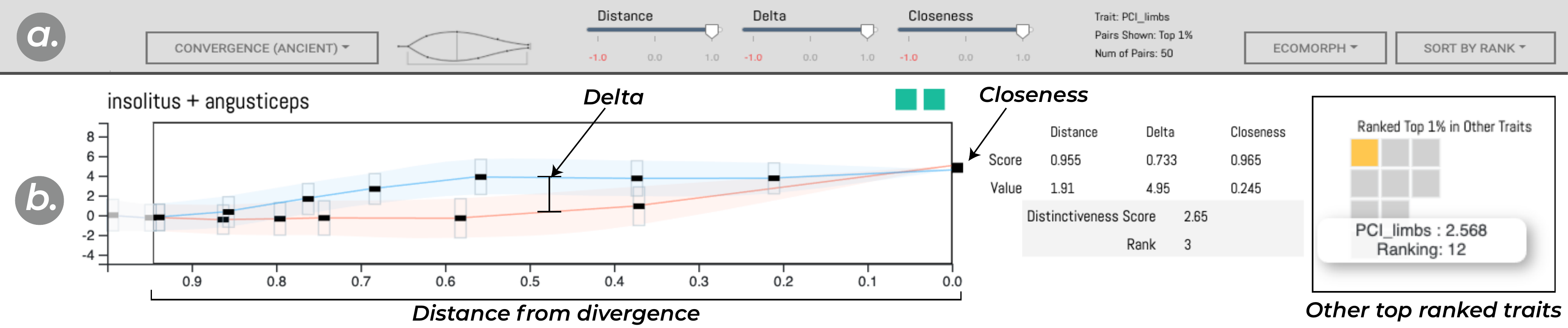}
 \vspace{-7mm}
 \caption{Pattern view components. (a) The user interface allows selection of a preset pattern, refined by adjusting the parameters for Distance, Delta, and Closeness. This interface also sorts rank pairs by top score or top rank frequency. (b) The first-ranked pair of paths (the species \textit{Anolis insolitus} and \textit{A. angusticeps}) for a convergence pattern for the trait ``snout vent length''. The line/area chart shows the most likely values and the associated uncertainty of the trait of consecutive species, with the ``delta'' between the species in the trait being evident in the middle. Individual species in the two extant species' ancestry are shown as rectangles. The heat map on the right show where other traits rank based on the selected pattern.}
  \vspace{-5mm}
 \label{fig:pair_plot}
\end{figure*}

The pattern view allows analysts to query for and find pairs of paths that follow a specific pattern of evolution such as convergence and divergence. Patterns of evolution are characterized by three key metrics: distance, delta, and closeness. The \textit{distance} between two species refers to time and topological distance up to the first common ancestor. \textit{Delta} is the maximum difference in an estimated continuous trait value after the species diverge. \textit{Closeness} is the difference in a specific, continuous trait value between the extant species. We developed a query interface, shown in Figure~\ref{fig:pair_plot}(a), that analysts can use to define patterns of interest based on these three characterizing parameters. We found that while these simple parameters cannot represent arbitrary patterns, they covered all the patterns of evolution our collaborators are interested in. 
To simplify the pattern definition, we also developed six preset patterns that an analyst can choose from to score pairs of paths. These patterns, shown in Figure~\ref{fig:pattern-breakdown}(b), emerged from repeated iterations with our collaborators [\trrrace{94}, \trrrace{96},\trrrace{129}].

To create a ranking for paths that match a specified pattern we calculate scores for all possible pairs of leaves using the selected pattern parameters for all traits. We then rank the pairs of paths based on the initial trait chosen by the domain expert, and visualize the two paths using a ranked list of line+area charts, as shown in Figure~\ref{fig:pair_plot}(b). In this chart, the vertical axis corresponds to the trait value. Individual species are shown as squares, which are positioned to be centered on their most likely trait value. The height of the box shows the confidence interval. The boxes are connected by lines for the most likely value, and areas for the confidence interval. 

One limitation of our original design of the pattern view was that it could only show a single trait at a time [\trrrace{96}]. In an early feedback meeting, our collaborators asked if it was possible to have an indication of whether a specific pair of paths was also ranked highly for other traits [\trrrace{99}, \trrrace{112}, \trrrace{113}]. That is, in some cases the analysts might be interested in identifying species pairs that have converged in several traits, rather than just one. Convergence of sets of traits is of particular interest because such cases can provide the strongest evidence for adaptation to particular environments. 
To address this shortcoming, we added a supplementary heat map to the side of the pair plot that indicates whether the pair is ranked in the top 1\% for a given pattern in any other traits in the data set, shown in Figure~\ref{fig:pair_plot}(b) on the right. Here, each square in a heat map represents other traits, where squares with darker saturation have a higher ranking. To find which pairs are ranked high for the pattern in the largest number of other traits, they can be sorted by frequency of top rankings from the heat map.

\subsection{Case Study}

We validate the trait and pattern views instantiated within Trevo by demonstrating their usefulness in a case study. The case study was conducted and written by our primary collaborators, who are also co-authors of this paper, and focuses on one of their primary datasets of the Anolis lizard genus. We provide a brief summary of findings here. We do not include the more detailed case study in this paper-within-a-paper, instead linking to it as external evidence [\trrrace{145}], as we find that domain-specific case studies often do not significantly contribute to a broader understanding of research contributions in design studies, but are rather akin to analysis scripts used in quantitative data analysis: they are necessary to ensure validity and trust, but do not convey knowledge on the subject of the research. 

Using the trait view, our collaborators were able to reduce their analysis to a subset of species that exhibit exceptionally large body features, and to see how body features evolved differently over time. Traditional visualization approaches would have required coloring disjunct branches in a phylogenetic tree and making difficult judgments about color variations; the trait view instead provided targeted analysis using spatial encoding of the traits of interest. With the pattern view, our collaborators were able to confirm a known convergence and divergence event, a task not possible with commonly used software for the phylogenetic analysis of trait evolution. Furthermore, they 
we were able to identify a new pattern of convergence in a pair of species, leading them to new biological questions about the evolutionary forces at play.
This case study shows that our collaborators not only could easily distinguish interesting patterns in their data using Trevo, but also document a previously unknown insight. We offer this case study as evidence of the validity of our proposed designs \cite{munzner_nested_2009}. 

\subsection{Conclusions}

We developed a web-based visualization tool we call Trevo in collaboration with evolutionary biologists to analyze phylogenetic trees with multivariate and uncertain attributes. In this paper-within-a-paper we contribute two novel visualization techniques implemented in the trait view and the pattern view. 
The two views prioritize visualizing the attributes in multivariate, phylogenetic trees over detailed topological information. This prioritization is by design. As the tree topology itself is highly uncertain, visualizing uncertain attributes on uncertain nodes is not helpful. Instead, our approach aggregates relevant subtrees by time and visualizes possible attribute distributions for temporal bins. 
The pattern view similarly prioritizes attributes with only rough topological measures, such as the time two species diverged. It is the first approach that allows researchers to query for complex evolutionary patterns based on a trait and topology, and explore these patterns across multiple traits. 

Trevo is being integrated into the computational workflow of the Harmon Lab. Additionally, the development of Trevo is part of a larger software project for creating MultiNet, a web-based tool for visualizing and analyzing multivariate networks~\cite{multinet}. The visualization insights generated from this design study are informing aspects of MultiNet's design. We discuss the methodological insights generated from this design study in the next section.

\section{Methodological Recommendations}
\label{sec:lessons-learned}
Our experiments with design study criteria for rigor --- specifically \crit{ABUNDANT}, \crit{REFLEXIVE}, and \crit{TRANSPARENT} --- offered us a wealth of opportunities to try new things, and to learn along the way. Through a critically reflective process, we distilled our learning into three methodological recommendations for conducting iterpretivist design studies.

\subsection{Explicit, systematic reflection is productive}
\label{sec:reflection-recommendation}

Reflection is a critical aspect of design study \cite{sedlmair_design_2012}, yet little is known on how and when to do this in practice \cite{2018_cga_reflection}. In our work we reflected regularly, and reflexively, documenting our reflections as we progressed through the study. What we found is that systematic reflection shifted the course of our research in productive and demonstrable ways.

For example, when Rogers arrived in our collaborators' lab at the start of the field study, she initially felt uncomfortable audio-recording her interactions with them. Because she was not familiar with the group and the group was not familiar with her, she felt like an intruder in the lab. In a reflective note from one of her first interviews, she noted: 
\begin{quote}
    \textit{I have not been recording these interviews as I am in the first week and I do not want to be intrusive.}
    [\trrrace{3}].
\end{quote}
She was, however, aware that audio recording would be beneficial to her note taking:
\begin{quote}
   \textit{I believe the recording will help me capture more than I can get from my note taking, and maybe more importantly, be more engaged in the interview process. I was initially hesitant to ask people to record them during my initial time here because I was new and unfamiliar and wanted our first interactions to be more candid.} [\trrrace{53}]. 
\end{quote}
The following week she decided to audio-record the participatory workshop she ran with the lab, and reflected on the experience:
\begin{quote}
    \textit{I am glad I recorded the workshop --- as I have re-listened to it and transcribed parts I felt were significant to the goal of the design study. Returning to the audio at a later time allowed me to notice things that people said when I was engaged in a conversation with someone else or did not have the base knowledge on a particular subject to want to write the moment down initially.} [\trrrace{36}]
\end{quote}

Rogers' concerns about her intrusive presence in the lab made her initially hesitant to audio-record interviews, to the detriment of her data gathering. After writing several reflexive memos detailing her feelings, and reflecting on the success of audio-recording the workshop, she changed her interview method and audio-recorded all interviews with collaborators. Off-loading the work of note-taking to the recording allowed her to engage in a more conversational, constructive way when conducting interviews:
\begin{quote}
    \textit{I found [audio-recording] extremely helpful as I was able to engage in conversation more easily than when I was attempting to take speed notes....The recording seems to blend into the scene and you forget it’s running after a couple of minutes. I will be using a recorder from now on.}[\trrrace{53}].
\end{quote}{}

By reflecting on her actions, Rogers was able to adjust and improve her research practices. Systematic, reflexive notes such as these are encouraged by qualitative researchers as they offer ``a partial means for providing checks on the researcher's own biases''  \cite{lincoln_establishing_1982} and a mechanism to  
``detect and correct deviations from the design goal early'' \cite{reymen_structured_2002}.

The start of audio-recording within the design study led to our second example of productive reflection. After conducting an interview, Rogers would listen to the audio-recording from the interview and reflectively transcribe it within a day or two. Transcription did not involve transcribing the audio-recording word for word, but was instead a reflective memo synthesizing the main points taken from the audio along with concrete quotes as evidence for these findings. When something stood out in the recording, Rogers would memo what time in the audio this happened, allowing her to easily revisit how something was said at a later time [\trrrace{36}]. 

We find that reflectively transcribing an audio-recording --- versus relying on an (automatically or externally generated) word-for-word transcription --- offers two advantages for analysis. First, listening to the audio while taking notes slows us down, allowing for a deeper, more thoughtful analysis. Writing down reflections requires us to stop, rewind, and listen to things multiple times, resulting in better notes and interpretations. Second, we find that listening to a recording allows us to re-experience the interview, but in a more detached and reflective way. This allows a new perspective on the discussion, separate from the one we experienced in the moment [\trrrace{126}]. 

Our third example of productive reflection occurred as we constructed an audit trail of our collected artifacts in order to produce a transparent trace of the design study process. Upon revisiting her old sketches, Rogers noticed that her design concepts for certain components were very narrow, particularly for an early version of the trait view [\trrrace{705}, \trrrace{709}, \trrrace{731}]. She reflected on the narrow design concepts during a meeting: 
\begin{quote}
    \textit{I get fixated on one design and I can see that in the sketches in my sketchbook.} [
    \trrrace{111}].
\end{quote}{}
This reflection prompted Rogers to attempt a redesign of the trait view's discrete plots, which had last gone through design iterations three months prior. Having recently reviewed her notes she took during the field study as she added them to the audit trail, she found new meaning, and new ideas for her redesign: 
\begin{quote}
    \textit{I still find details that I missed at the time of a meeting or at an initial reflection.} [\trrrace{111}].
\end{quote}{}
The redesigned trait view, shown in Figure~\ref{fig:trait_view}, shows relationships between trait values and their probability distributions that were not shown in earlier designs.

We did not anticipate that the act of curating and organizing artifacts would facilitate productive reflection and play a role in design development. 
This redesign would likely not have occurred without the reflective processes of revisiting past notes, a concept emphasized in work on systematic reflection for design in engineering. By adopting regular reflection during design,
``the chance of overlooking important aspects is decreased'' \cite{reymen_structured_2002}.
Tavory and Timmermans advocate for revisiting experiential notes to reconsider them with newfound knowledge or perspective: ``We are constantly re-experiencing parts of our world as we go about the business of living. When we move through our surroundings, we not only encounter new problem situations but find new problems in old situations'' \cite{tavory_abductive_2014}. 

\vspace{0.25cm}
\noindent \textbf{RECOMMENDATION} Our work shows that adopting regular, systematic, reflective practices within a design study can improve the research methods, domain understanding, and visualization designs. We recommend four opportunities for reflection-on-action. First, take reflective notes before and after interviews with domain collaborators. This activity takes only a few minutes but significantly improves the focus of an interview as well as captures initial interpretations and ideas for next steps. Second, include reflexive considerations in your field notes. Reflecting on changing perspectives, biases, methodological rationale, and feelings can be a valuable source of insight.
Third, audio-record interviews and analyze them via reflective transcription. The reflective transcription should occur soon after the interview to support experiential recall on the part of the researcher. Fourth, revisit early notes and sketches. During these revisits look for opportunities to reinterpret experiences through a new lens of deeper understanding.

\subsection{Traceability supports transparency and reflection}
\label{sec:trrrace}

Providing a transparent, scrutinizable trace of a design study is essential for allowing judgments about the quality of the research \cite{meyer_criteria_2019}. As we developed an auditable trace through our collection of research artifacts, we found, however, that revisiting evidence \textit{also} supported productive reflection that shifted and changed the course of the study. Supporting different ways to trace the design study process was important for encouraging both transparency and reflection in our study.

From the start of the field study, we meticulously collected a rich set of research artifacts in order to abundantly document our research process. We stored the artifacts in an online repository, and created a record for each in a spreadsheet that included a descriptive title, the date it was created, a unique id, and the research artifact \textit{type} such as meeting note, sketch, email, etc. Building on this collection of evidence, we experimented with transparent reporting by creating an audit trail of the artifacts. Our initial, web-based design of the audit trail was inspired by those created for other experiments on reporting design studies \cite{kerzner_framework_2019,mccurdy_framework_2019}. Like previous examples, our website traced the design study temporally by visually organizing artifacts on an overview timeline, and providing access to the recorded artifacts themselves through a details-on-demand side panel. Each artifact is represented on the timeline as a square, color-coded by its type 
[\trrrace{161}].

While building the audit trail, we reflectively engaged with the research artifacts, leading to demonstrable changes within our study, as we previously discussed in Section \ref{sec:reflection-recommendation}. This engagement shifted the audit trail toward use as an internal, research tool. We found that we wanted to trace research \textit{concepts} across the study, including our growing understanding of domain principles such as convergence and uncertainty, as well as our criteria experiments through reflexivity and sketching. To support concept tracing we extended our metadata for each research artifact to include tags that pull information embedded within the artifacts. These concept tags allow for a trace of how our awareness and understanding of various concepts evolved throughout the study. We extended the website to include the concept tags for each artifact in the detail view; clicking on a specific tag highlights other artifacts with the same tag in the timeline overview
[\trrrace{161}].

The final iteration of our tool supports an unanticipated range of research tasks: recording diverse research artifacts, reflecting on conceptual developments, and reporting on the design study process. It is a trace of our research process from two perspectives: a temporal perspective for transparent and auditable reporting and a conceptual perspective for reflective research practices. We consider this tool to be a \textbf{trrrace}, as both a speculative nod to material traces \cite{offenhuber_data_2019} and to the record, reflect, and report tasks it supports.

The trrrace has theoretical connections to both audit trails \cite{lincoln_establishing_1982,akkerman2008auditing} and annotated portfolios \cite{gaverbill_annotated_2012,bowers_logic_2012}. 
As referential material \cite{lincoln_establishing_1982}, our research artifacts are evidence of the design study process \cite{wacharamanotham_transparency_2020, middleton_recommendations_2018, hartmann_pictionaire:_2010, zimmerman_role_2008}, capturing fleeting aspects of the study that led to insight. Organizing these artifacts temporally provides a trace of the study itself \cite{offenhuber_data_2019}, providing an auditable mechanism for reviewing the quality of the research \cite{talkadsukumar_transparency_2020, wacharamanotham_transparency_2020, moravcsik_transparency:_2014}. 
Our research artifacts are also manifestations of design knowledge \cite{gaverbill_annotated_2012}, with the knowledge engrained within the artifact \cite{cross2004expertise}. Each artifact's concept tag, created from the artifact itself, is an annotation, allowing for a trace that connects seemingly disparate artifacts through more general concepts.
These theoretical connections point to an opportunity for further theorizing about, and experimenting with, design study trrraces.

\vspace{0.25cm}
\noindent \textbf{RECOMMENDATION } Our experiments with abundant evidence and transparent reporting led us to the concept of a trrrace, which supports recording, reflecting, and reporting in design study. We recommend that design study researchers plan for a trrrace early in a study and consider three important issues. First, the process of collecting artifacts greatly benefits from establishing a system for organization early on. We used an online spreadsheet and adopted a regular practice of adding records of digitized artifacts as we generated them. Second, develop mechanisms to automatically extract concept tags from the artifacts themselves. We extracted concepts from the artifacts manually for this project, but in future work we plan to develop an improved, semi-automated approach. 
Third, the immersive, ethnographic nature of design study requires considerations of how to handle privacy, as well as anonymization for review. We encourage developing a system for anonymizing artifacts early in the study process. Additionally, we find that the best method for navigating transparent recording of a study is to be transparent: tell your collaborators when you are recording, establish what will be on- versus off-record, provide them access to your notes, and be aware of recording delicate social dynamics. 

\subsection{Methods are necessary, but evidence is the proof}

Employing appropriate and justified research methods within a design study is necessary for achieving rigor, but a checklist of methods is not sufficient for arguing that a study is rigorous. 
The design study rigor criteria are meant to provide guidance on \textit{what} to achieve, not \textit{how} to do so \cite{meyer_criteria_2019}.
Evidence of the criteria within a study
is the proof. The type, extent, and depth of evidence that is sufficient for arguing that a design study meets the criteria for rigor, however, is an open question, and likely one without a standardizable answer. As part of our experiments we reflected over our research artifacts and experiences, looking for evidence of the criteria. We found that shifts in the way we communicated and interacted with our collaborators suggest that our study was \crit{INFORMED} and \crit{ABUNDANT}. 
 
During the early stages of our field study, work discussions with our collaborators centered around semistructured interviews. We organized interviews to have 2-3 in a single day and scheduled interview days every few days. Rogers saved up questions she had until these interviews [\trrrace{7}, \trrrace{53}]. The infrequent discussions were relatively long in duration, lasting from 1-2 hours at a time, were dense with domain information, and had a formal tone. Post-interviews, Rogers would revisit and look up domain concepts and vocabulary that emerged from these interviews as she was building her understanding of the domain: 
\begin{quote}
    \textit{The paper [linked] below was a really good resource for getting an understanding of the group comparisons that indicate adaptive events such as convergence.... People I have interviewed touched on these concepts, but because the concepts are complex and varied, it is really hard to get a good synthesis of the main points. I feel as if I am hearing recurring words that come up in conversation, but I have been missing the connection between them.} [\trrrace{87}]
\end{quote}

As the field study progressed and Rogers felt increasingly comfortable asking questions outside of scheduled interviews, work communications shifted to shorter, informal discussions and texts. The language of communications also shifted as Rogers increasingly used domain vocabulary and concepts fluently. For example, Rogers saw some unexpected biological relationships in the data while developing one of the visualization views, and messaged a collaborator to confirm her observations:
\begin{quote}
    \textit{Rogers: WOOP. Saving the summary view.\\
    AP: Booyah\\
    Rogers: One thing I noticed the other night is that a lot of the convergent pairs are not both the same ecomorph --- but because we are looking at a single trait, would it make sense that two ecomorphs would have similar characteristics for a single trait? Ex: trunk crown and twig having similar PCIII Padwitch vs tail?!\\
    AP: Hmmmm.... You're finding that even when using the PC traits? Because those PCs are essentially composites of multiple traits [\trrrace{133}]\\}
\end{quote}
\vspace{-3mm}
The texts continued as Rogers also excitedly communicated her findings of a problem with the pattern ranking system:
\begin{quote}
    \textit{Rogers: THINK I FOUND A BUG IN THE DELTA.\\
    AP: Oh *** What's it doing? [\trrrace{133}]\\}
\end{quote}
\vspace{-3mm}
At the time of this text exchange, Rogers had spent significant time engaged with the domain, and she understood enough about domain concepts to identify mismatches in what she saw in the data. Furthermore, identifying these mismatches \textit{excited} her.

This exchange aligns with indicators for immersion: using domain-specific language the researcher engages in ``informal peer-to-peer communication with domain experts about domain science and visualizations'', eventually becoming ``concerned with, affected by, and personally involved in the other domain''  \cite{hall_design_2019}. Design by immersion is an approach that, through long-term, committed engagement, provides visualization researchers an abundant exposure to a domain space, allowing them to develop a deeply informed understanding. 

Every design study, like other qualitative inquires, is unique in complex ways and thus requires the construction of careful, thoughtful arguments for its quality: ``Excellent research is not achieved solely by the use of appropriate strategies or techniques. The skillful use of strategies only sets the stage for the conduct of inquiry'' \cite{morse2020changing}. 
Changes in the way we communicated with our collaborators --- not the time we spent in the field or the number of interviews we conducted ---  suggests that our design study met aspects of both the \crit{INFORMED} and \crit{ABUNDANT} criteria. We argue for careful argumentation, backed up by rich evidence and grounded in existing literature and theories, as a general model for supporting claims of rigor in design studies. Being reflexive and noticing not only how we affected the research, but also how it affected us, offered us opportunities to more deeply reflect on the impacts of our criteria experimentations. We speculate that many such opportunities may be found in any design study.

\vspace{0.25cm}
\noindent \textbf{RECOMMENDATION } Knowing when a design study has reached a critical threshold for establishing rigor is difficult, with no single, objective metric. Through critical reflection we positioned our experiences and evidence --- shifting patterns of communication --- within existing theoretical concepts --- design by immersion \cite{hall_design_2019} --- allowing us to build links between what occurred in our research and what it could mean. We recommend that design study researchers plan for the time and space to engage critically and reflectively with their research artifacts and experiences; propose, repropose, and repropose again how what they learned engages with the existing literature; and resist the urge to argue that a study is rigorous because of a checklist of methods employed, instead looking for things that changed, shifted, and surprised. 

\section{Conclusion}
\label{sec:conclusion}

This paper reports on an interpretivist design study and a resulting diverse set of knowledge outcomes consisting of visualization techniques, methodological insights, and new methods for reporting. We found that our experiments with establishing rigor through the \crit{ABUNDANT}, \crit{REFLEXIVE}, and \crit{TRANSPARENT} criteria led to a myriad of learning opportunities. Yet those opportunities are messy, overlapping, and difficult to distill. For example, our efforts to provide transparency relied on abundant data collection, and (reflexively) changed our writing methods as we crafted this paper. We learned much more than we have reported, but the challenge of aligning the evidence, our experience, and existing theory kept us from fully synthesizing the rich learning this interpretivist design study provided.

One such example is the trrrace construct we propose for recording, reflecting, and reporting in design study. The idea of the trrrace emerged as we worked to enhance the transparency of this report. The more we linked into our collection of artifacts, the more we noticed how these links provided useful traces of our research process. We also became aware of challenges for a mechanism like a trrrace that is used in both the research and reporting processes: how do we ensure persistence of the trrrace and the myriad artifacts it links together? How do we consider privacy concerns, as well as anonymization constraints? How do we develop and maintain a trrrace in a way that does not slow down design-oriented research? How do we improve our recording practices to enhance the traceability of a trrrace? 
How do we report a trrrace in a way that is accessible, understandable, and scrutinizable?

This last question offers opportunities to reflect on current practices for reporting through traditional supplemental materials that can, at best, tell a curated story parallel to a paper, but at worst, can be an impenetrable dump of information. What types of visualizations, interactions, and interfaces can we design to help a reader navigate a trrrace? How might we tell a data-driven story from an abundant collection of evidence? If we embrace the concept of material traces, how might this fundamentally change the way we think about supplemental materials, transparency, and reproducability?
Developing theory and pragmatic guidance for design study trrraces is one of the more exciting future directions pointed to by this work. We hope this paper is a catalyst for further conversations about trrraces, as well as the broader opportunities and challenges for interpretivist design study.

\acknowledgments{We thank the members of the Harmon lab, the Visualization Design Lab, and the MultiNet team for their participation, feedback, and support. We also appreciate the thoughtful comments from the anonymous reviewers that helped to strengthen this paper. We gratefully acknowledge funding by the National Science Foundation (OAC 1835904). 
}

\bibliographystyle{abbrv-doi}

\bibliography{2020_trevo.bib,mm-bib}
\end{document}